# ENHANCED SENSING CHARACTERISTICS IN MEMS-BASED FORMALDEHYDE GAS SENSOR

*Yu-Hsiang Wang[1], Ching-Cheng Hsiao[1], Chia-Yen Lee[1], Rong-Hua Ma[2], Po-Cheng Chou[3]*

[1]Department of Mechanical and Automation Engineering, Da-Yeh University, Changhua, Taiwan,
[2]Department of Mechanical Engineering, Chinese Military Academy, Kaohsiung, Taiwan,
[3]Department of Interior Design, Shu-Te University of Science and Technology, Kaohsiung, Taiwan.

## ABSTRACT

In this study, the proposed sensor integrates a sensing layer, a heating device, and electrodes on the substrate. The micro heater is integrated in the sensor to provide instantaneous and precise temperature control capability. The electrodes are fabricated to connect resistance meter for measuring variation of electrical conductibility of the sensing layer. The grain size of the NiO thin film is almost to be nanometer level, and therefore both the sensitivity and the lowest sensing limit of the device are enhanced due to the enlarged area of the catalyst grains contacting with the surrounding gas. The experimental data show that decreasing thickness of sensing layer in the sputtering process significantly increases the sensitivities of the gas sensor and improves its lowest detection limit capability (0.7 ppm). Although we can further improve lowest detection limit by co-sputtering with $NiO/Al_2O_3$ (40 ppb), it needs to consider that selectivity will be reduced. The integrated micro heater simplifies the experimental set-up and can be realized using a simple fabrication process. The presented microfabricated formaldehyde gas sensor with a self-heating $NiO/Al_2O_3$ thin film is suitable not only for industrial process monitoring but for safeguarding the human health in buildings.

## 1. INTRODUCTION

Formaldehyde is an important chemical that used widely by industry to manufacture building materials and numerous household products. But it can cause irritation of the skin, eyes, nose, and throat. High levels of exposure may cause some types of cancers. It has been shown that throat and nose irritation can occur at formaldehyde levels as low as 0.08 ppm [1]. The WHO (World Health Organization) has established a permissible long-term exposure limit of 0.1 ppm [2]. The methods of detection for formaldehyde gas may divide into three main categories: GC/MS, optical, and MEMS based gas sensor. The principle of GC/MS is achieved using a stringent semi-batch air-sampling procedure, followed by a batch analysis of the sample, but it is benchtop and sampled by a tube with absorbent. Thus numerous researchers have studied about optical sensor for formaldehyde quantification applications [3-4]. Even though the optical sensors are simultaneous sampling and have instantaneous analyzing time, associated optical arrangements tend to be rather bulky and elaborate. In the last decade, emerging MEMS and micromachining techniques have led to the development of miniaturized sensing instrumentation capable of accessing information at a microscale level. Importantly, the functionality and reliability of these micro sensors can be increased through their integration with mature logic IC technology or with other sensors.

Recently, Dirksen et al. [2] proposed a NiO thin-film formaldehyde gas sensors by dipping alumina substrates in a nickel acetylacetoneate solution to form thin NiO films of thickness 0.5 μm. It was found that the conductivity of these films changed as the formaldehyde concentration was varied at temperatures ranging from 400-600 °C. At approximately 3 μm, the sintered grain size was rather large and was hence expected to reduce the sensitivity of the device. However, a linear formaldehyde sensitivity of 0.825 mV ppm$^{-1}$ was attained at a temperature of 600 °C. Although the sensor demonstrated a high sensitivity, its detection limit was only 50 ppm, i.e. far higher than the prescribed "maximum permissible long-term exposure" limit of 0.1 ppm. Furthermore, the proposed sensor lacked an integrated heating device capable of maintaining the optimal working temperature of 600 °C. Therefore, the sensor operation required the use of an external heater, which not only increased the bulk of the sensor arrangement, but also increased its power consumption. As a means of overcoming this problem, Lee et al. [5] proposed the use of Pt resistors as integrated micro-heaters for MEMS-based temperature control systems. Pt was chosen specifically as the resistor material on account of its physical and chemical stability. The current study develops a new process for the fabrication of a MEMS-based formaldehyde sensing device comprising micro heater and electrodes with Pt resistance heaters and a sputtered NiO/Al$_2$O$_3$ layer. The experiment data indicate a high sensitivity, a low detection limit, a simple arrangement with no requirement for any from of external heating device.

## 2. EXPERIMENTAL

### 2.1. Design

In this study, the proposed sensor integrates a sensing layer, a heating device, and electrodes on the substrate. The micro heater is integrated in the sensor to provide instantaneous and precise temperature control capability. The electrodes are fabricated to connect resistance meter for measuring variation of electrical conductibility of the sensing layer. The grain size of the NiO thin film is almost to be nanometer level, and therefore both the sensitivity and the lowest sensing limit of the device are enhanced due to the enlarged area of the catalyst grains contacting with the surrounding gas. As shown in figure 1, we design two types of sensors for different micro heater positions and sensing layers. Type A is the micro heater above the sensing layer, and the other one (type B) is fabricated the micro heater between the sensing layer and the substrate. Table 1 shows the detail of design in this study, we not only tried to compare the characteristic of sensor with type A and type B but also attempted decrease the thickness of sensing layer and deposited the sensing layer by "co-sputtering" with NiO/Al$_2$O$_3$ to improve the sensitivity and the lowest sensing limit of the sensor.

### 2.2. Microfabrication

The gas sensors of this study were fabricated by using quartz substrates as the substrate. Figure 2 presents a schematic illustration of the fabrication process of type A and type B. As shown in figure 2(a) (type A) a NiO sensing layer was prepared using an RF magnetron sputtering system with a NiO target of 99.98% purity [8]. The oxide was sputtered on substrates, which were placed at a distance of 11.4 cm from the NiO target. Sputtering was performed under a gas pressure of 0.01 torr with the target maintained at a constant RF power of 200 W. The reactive sputter gas was a mixture of argon (50%) and pure oxygen gas (50%). The substrate temperature during sputtering was 400 $^{o}$C [9]. Prior to deposition, the chamber was pumped to a background pressure of 10-6 torr for 1 h and a pre-sputtering process was performed for 10 min to clean the target surface by removing any possible traces of contamination. To control the thin film thickness, the deposition time was 10 hours. The deposited NiO films was found to have thickness of approximately 1500 Å. A thin layer of Cr (0.02 μm) was then deposited as an adhesion layer for the subsequent electro-beam evaporation of a Pt layer of 0.2 μm thickness. A standard lift-off process was employed to pattern the Pt/Cr layer to form a micro heater and the electrodes on the NiO sensing layer. The resistance of the heater was designed to be 30 Ω. As shown as figure 2(b) (type B), electron-beam evaporation and lift-off procedures were then repeated to deposit and shape Pt micro heater and electrodes. Then NiO sensing layer was sputtering-deposited.

Table 1: Design of formaldehyde sensors with different micro heater position and parameters.

| Micro heater position<br>Sensing layer thickness | Above sensing layer | Below sensing layer |
|---|---|---|
| 1500Å | A$_{1500}$ | B$_{1500}$ |
| 500Å |  | B$_{500-NiO}$ |
| 500Å (co-sputtering) |  | B$_{500-NiO/Al2O3}$ |

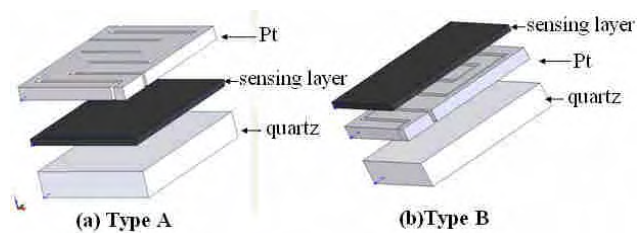

Fig. 1: Schematic illustration of formaldehyde sensors with different micro heater position.

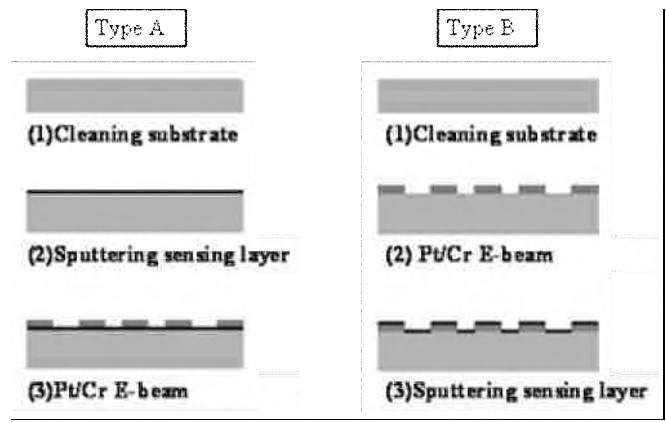

Fig. 2: schematic illustration of the simplified fabrication process of type A and type B

## 3. RESULTS AND DISCUSSION

### 3.1. Effect of substrate temperature and micro heater position with sensing layer

As shown in figure 3, a linear dependency is observed between the resistance and the formaldehyde concentration at different sensor working temperatures (i.e. micro-heater temperatures). The slopes of the plotted lines represent the sensitivity of the device and are found to be -0.137 $\Omega$ ppm$^{-1}$ at 280 $^o$C, -0.12 $\Omega$ ppm$^{-1}$ at 215 $^o$C, and -0.104 $\Omega$ ppm$^{-1}$ at 150 $^o$C, respectively. The lowest detection limit of the type $A_{1500}$ sensor is determined to be 1.2 ppm at 280 $^o$C. As shown in figure 4, a linear dependency is also observed between the resistance and the formaldehyde concentration at different sensor working temperatures. The sensitivities of the type $B_{1500}$ are found to be -0.335 $\Omega$ ppm$^{-1}$ at 300 $^o$C, -0.293 $\Omega$ ppm$^{-1}$ at 250 $^o$C, and -0.181 $\Omega$ ppm$^{-1}$ at 180 $^o$C, respectively. The lowest detection limit of the type $B_{1500}$ sensor is determined to be 0.8 ppm at 280 $^o$C By the results between type $A_{1500}$ and $B_{1500}$, it is clear to find:

1. When the working temperatures of the substrate increases, the sensitivity of the sensor increases.

2. Because the area of the sensing layer contacting with the surrounding gas on sensor type B is larger than that of type A, the sensitivity level increases and the lowest detection limit is also improved from 1.20 ppm to 0.8 ppm.

### 3.2. Effect of sensing layer thickness and materials

For approaching higher sensitivity and better characteristics of the sensors, we fabricated the sensing layer thickness approximately 500 Å by type B fabrication, and deposited the sensing layer by "co-sputtering" with NiO/Al$_2$O$_3$ to compare the results at different sensing layer thickness and materials of sensing layer. Figure 5 and 6 show X-ray diffraction analysis (XRD) for sensing layer of type $B_{500\text{-NiO}}$ and type $B_{500\text{-NiO/Al2O3}}$, respectively. In figure 5 and 6, we can clearly find the peak of NiO and Pt. However, we cannot find any peak of Al$_2$O$_3$ in figure 6, even though we deposited the sensing layer of $B_{500\text{-NiO/Al2O3}}$ by co-sputtering technology with NiO/Al$_2$O$_3$. The Al$_2$O$_3$ of the sensing layer of $B_{500\text{-NiO/Al2O3}}$ is amorphous alumina [7]. In figure 7 and 8, the lowest detection limit of type $B_{500\text{-NiO}}$ and $B_{500\text{-NiO/Al2O3}}$ were determined, respectively. We can find the sensing layer of NiO/Al$_2$O$_3$ improve the lowest detection limit from 0.7ppm to 40ppb. Although it can improve the lowest detection limit by co-sputtering with NiO/Al$_2$O$_3$, but the selectivity of the sensor will also be reduced, as shown in figure 9 and 10. In figure 9, a high selectivity over a wide formaldehyde concentration range in the presence of interfering species such as acetone, ethanol and methanol. However, figure 10 indicates the selectivity of type $B_{500\text{-NiO/Al2O3}}$ was reduced by sensing layer of NiO/Al$_2$O$_3$.

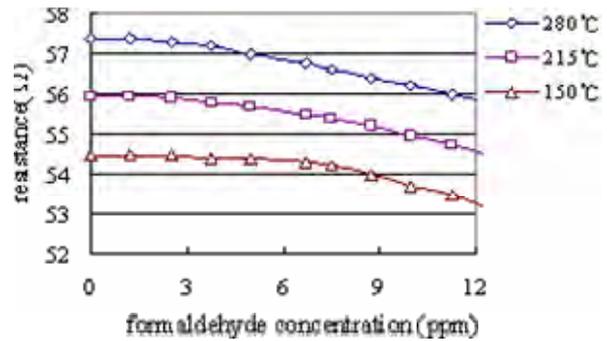

Fig. 3: Experimental results of type $A_{1500}$ sensor at different work temperature.

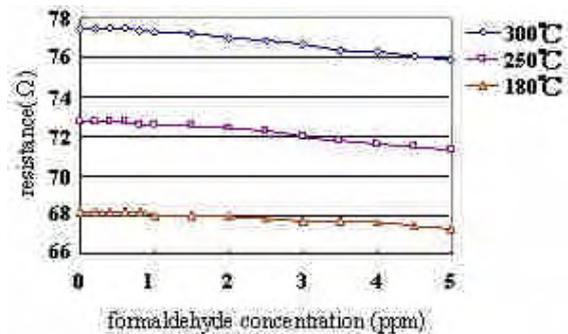

Fig. 4: Experimental results of type $B_{1500}$ sensor at different work temperature.

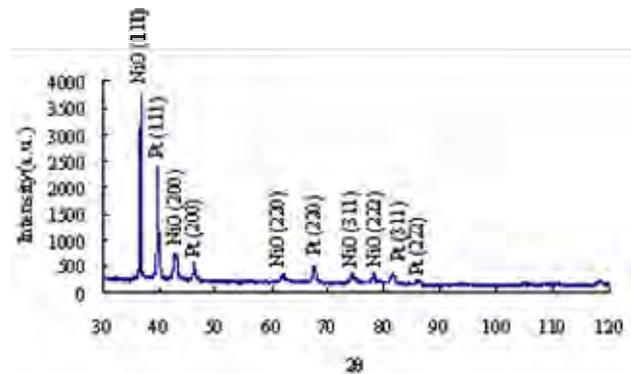

Fig. 5: XRD diffraction patterns of NiO thin film.

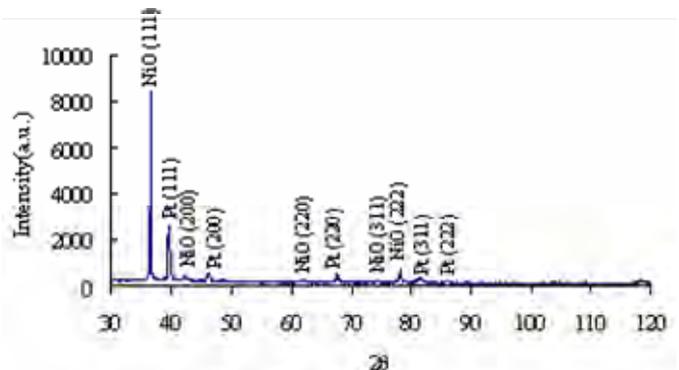

Fig. 6: XRD diffraction patterns of NiO/Al2O3 thin film.

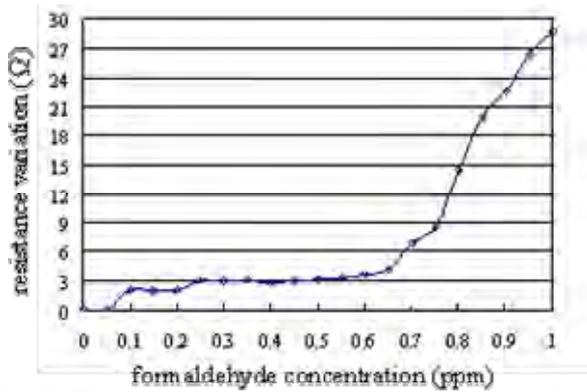

Fig. 7: The lowest detection limit of the sensor of type $B_{500\text{-NiO}}$.

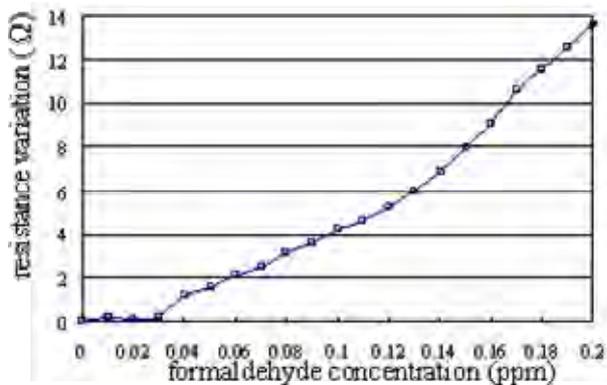

Fig. 8: The lowest detection limit of the sensor of type $B_{500\text{-NiO/Al2O3}}$.

### 3.3. Time response

In conventional gas detectors, the time required for formaldehyde concentration measurement can vary from hours to days. However, a requirement exists for sensors with a real-time gas detection and measurement capability. Figure 11 and 12 presents the time response of the gas sensors developed in the present study. The average time constant of the proposed formaldehyde gas sensor is determined to be 7 s and 6 s for type $B_{500\text{-NiO}}$ and $B_{500\text{-NiO/Al2O3}}$ sensor at a micro-heater temperature of 300 $^{\circ}$C.

### 4. CONCLUSIONS

This study has successfully demonstrated a novel self-heating formaldehyde gas sensor based on a thin film of NiO sensing layer. A new fabrication process has been developed in which the Pt micro-heater and electrodes are deposited directly on the substrate and the NiO thin film is deposited above on the micro heater to serve as sensing layer. Pt electrodes are formed below the sensing layer to measure the electrical conductivity changes caused by formaldehyde oxidation at the oxide surface. The experimental data show that decreasing thickness of sensing layer in the sputtering process significantly increases the sensitivities of the gas sensor and improves its lowest detection limit capability (0.7 ppm). Although we can further improve lowest detection limit by co-sputtering with $NiO/Al_2O_3$ (40 ppb), it needs to consider that selectivity will be reduced. The integrated micro heater simplifies the experimental set-up and can be realized using a simple fabrication process. The presented microfabricated formaldehyde gas sensor with a self-heating $NiO/Al_2O_3$ thin film is suitable not only for industrial process monitoring but for safeguarding the human health in buildings.

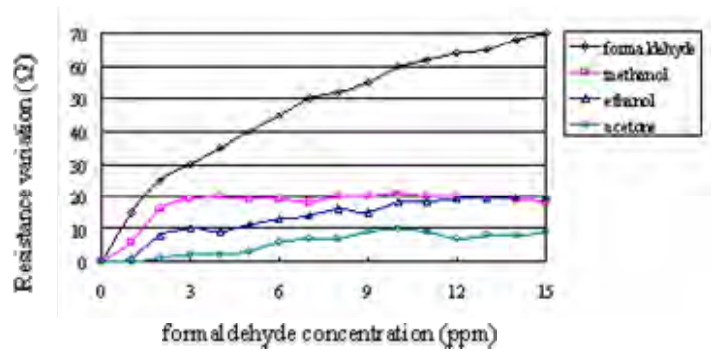

Fig. 9: The selectivity of the sensor type $B_{500\text{-NiO}}$.

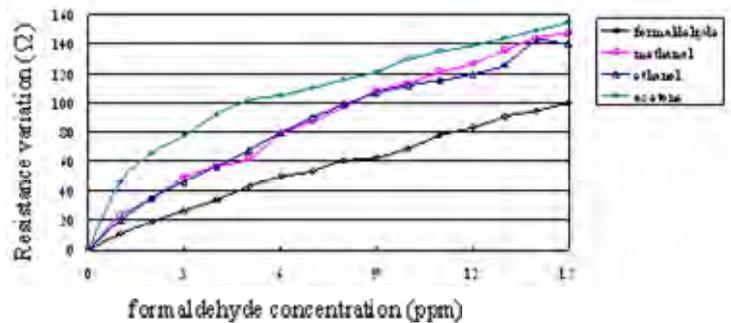

Fig. 10: The selectivity of the sensor type $B_{500\text{-NiO/Al2O3}}$.

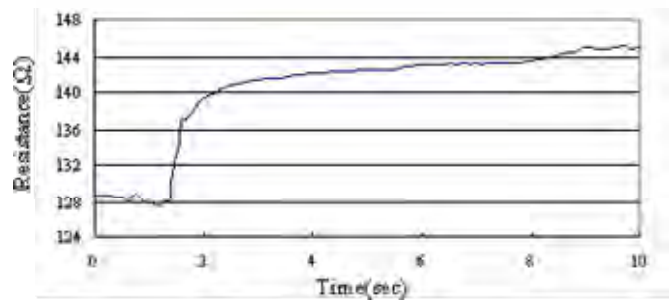

Fig. 11: Response transient of the formaldehyde gas sensor for type $B_{500\text{-NiO}}$.

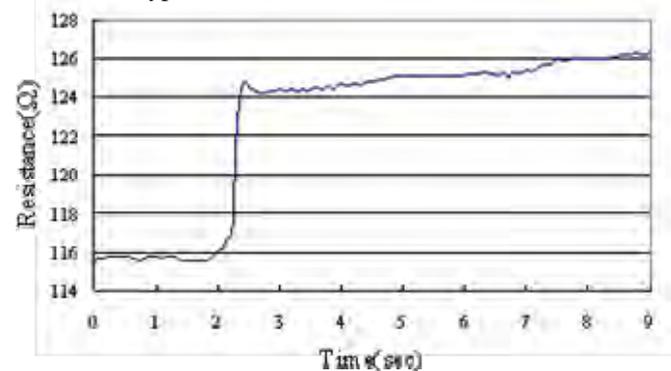

Fig. 12: Response transient of the formaldehyde gas sensor for type $B_{500\text{-NiO/Al2O3}}$.

## ACKNOWLEDGEMENTS

The authors would like to thank the financial support provided by the National Science Council in Taiwan (NSC 95-2211-E-212-058 and NSC 95-2218-E-006-022).